\documentclass[aps,prl,twocolumn,amsmath,amssymb,nofootinbib,superscriptaddress,longbibliography]{revtex4-2}

\usepackage{xcolor}
\usepackage{soul}
\usepackage{physics}
\usepackage{graphicx}
\usepackage{amsmath}
\usepackage{amssymb}
\usepackage{xcolor}
\usepackage[english]{babel}

\usepackage{hyperref}
\hypersetup{pdfstartview={FitH},pdfpagemode={UseNone},
            colorlinks,linkcolor=blue, citecolor=blue, urlcolor=blue,
            bookmarksopen=true, pdfnewwindow=true}
\usepackage[all]{hypcap}

\begin{document}

\title{Demonstration of Lossy Linear Transformations and\\Two-Photon Interference on a Photonic Chip}

\author{Kai Wang}

\thanks{these authors contributed equally to this work}

\affiliation{Department of Physics, McGill University, 3600 rue University, Montreal, Quebec H3A 2T8, Canada}

\author{Simon J. U. White}

\thanks{these authors contributed equally to this work}

\affiliation{Centre for Quantum Dynamics and Centre for Quantum Computation and Communication Technology (CQC\textsuperscript{2}T), Griffith University, Yuggera Country, Brisbane, 4111 Australia}

\affiliation{School of Mathematical and Physical Sciences, University of Technology Sydney, Ultimo, NSW, 2007, Australia}

\author{Alexander Szameit}

\affiliation{Institut f{\"{u}}r Physik, Universit{\"{a}}t Rostock, Albert-Einstein-Stra{\ss}e 23, 18059 Rostock, Germany}

\author{Andrey A. Sukhorukov}

\affiliation{Centre of Excellence for Transformative Meta-Optical Systems (TMOS), Department of Electronic Materials Engineering,  Research School of Physics, Australian National University, Canberra, ACT 2600, Australia}

\author{Alexander S. Solntsev}

\email{alexander.solntsev@uts.edu.au}

\affiliation{School of Mathematical and Physical Sciences, University of Technology Sydney, Ultimo, NSW, 2007, Australia}

\begin{abstract}
Studying quantum correlations in the presence of loss is of critical importance for the physical modeling of real quantum systems. Here, we demonstrate the control of spatial correlations between entangled photons in a photonic chip, designed and modeled using the singular value decomposition approach. We show that engineered loss, using an auxiliary waveguide, allows one to invert the spatial statistics from bunching to antibunching. Furthermore, we study the photon statistics within the loss-emulating channel and observe photon coincidences, which may provide insights into the design of quantum photonic integrated chips.
\end{abstract} 

\maketitle

Photonic quantum information processing is poised to revolutionize a range of areas in computation and telecommunication, enabling superior approaches to cryptography, random number generation, numerical modeling, and data processing~\cite{Knill2001, OBrien2009, Aspuru-Guzik2012, Walmsley525, Wang285, Solntsev2021}. One of the key phenomena enabling these advances is photon entanglement combined with the underlying quantum interference. Various experiments have been conducted to demonstrate the potential to manipulate photon correlations, observing the transition between spatial bunching and antibunching in a range of systems, such as bulk optics, plasmonics, and metasurfaces~\cite{Hong1987, PhysRevA.58.4904, Zeilinger1981, Vest1373, Wang2018,Li2021}. This interesting phenomenon can also be perceived through the lens of switching the photon statistics from traditional bosonic to fermionic-like behavior~\cite{PhysRevA.53.R1209, PhysRevA.45.7729, PhysRevLett.108.010502, Matthews2013}.

Photonic integrated circuits have emerged as an extremely promising platform for the implementation of quantum optical applications due to their miniaturized nature and excellent scalability~\cite{SOLNTSEV201719, Shen2017, PhysRevA.92.033815, Wang285, Wang2019,Klauck2019,White_2020}. In these circuits, photon state transformations can be controlled with high precision and stability, including non-unitary approaches that utilize loss ~\cite{Carolan711, Clements:16, Bell:2021-70804:APLP, Ehrhardt2022}. Recently a universal method to treat non-unitary transformations has been proposed~\cite{Miller:12, Shen2017, PhysRevX.8.021017}, based on singular value decomposition (SVD); however, experimental implementations of this approach remained limited to classical light. 

In this work, we show an experimental demonstration of a lossy beamsplitter, modeled using SVD, enabling the control of non-classical statistics by studying the dynamics of photon pairs in a system of coupled waveguides. We demonstrate that, with appropriate design, one can achieve full control of photon-pair correlations, generalizing the results previously reported for lossy beamsplitters and metasurfaces~\cite{Vest1373, Wang2018,Li2021, Lung2020}. Crucially, this work also addresses the question of what happens with photons and their correlations in the channel that emulates loss, which may offer insight for the design of large-scale quantum photonic circuits relying on beam splitters and photon interference~\cite{Shen2017}.

\begin{figure}[t]
\centering
\includegraphics[width=\linewidth]{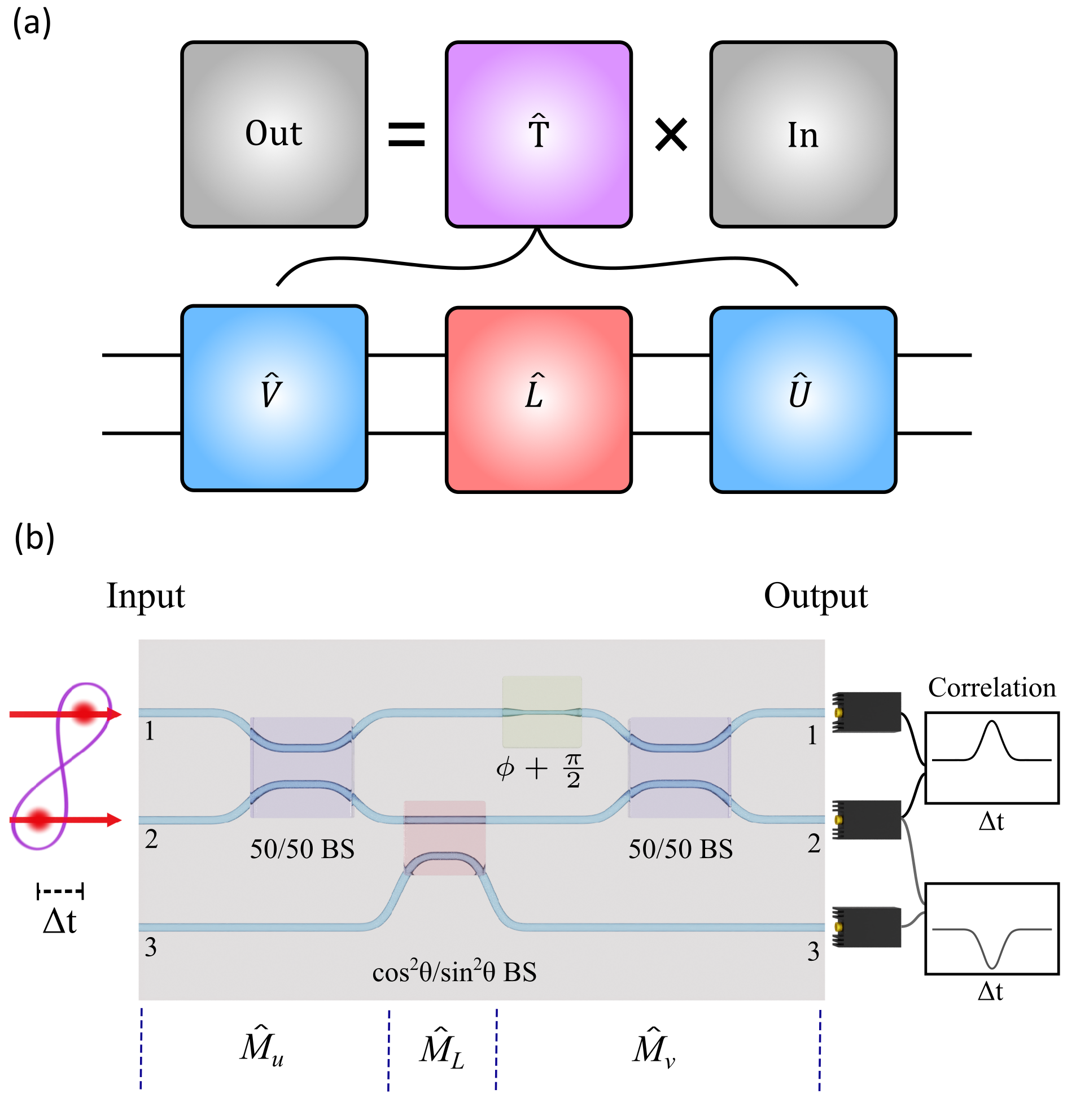}
\caption{Model and implementation of non-unitary transformations (a) Arbitrary $2\times2$ transformations (T) can be computed using a combination of unitary matrices ($\hat{V}$ and $\hat{U}$) and diagonal matrices ($\hat{L}$). (b) Conceptual diagram for implementation of a non-Hermitian Mach-Zehnder interferometer. The Hermitian couplers and the phase shifter are highlighted in blue and yellow respectively, and loss from a single mode is realized through coupling to an ancillary mode highlighted in red. The lossy $2\times2$ transformation, $\hat{T}$, only considers ports 1 and 2, whereas a full description including the ancillary port is given by the $3\times3$ matrices $\hat{M}_U$, $\hat{M}_L$, and $\hat{M}_V$. The diagram is not shown to scale.}
\label{fig:1}
\end{figure}

We begin by considering a general $2\times2$ linear transformation $\hat{T}$. As conceptually sketched in Fig.~\ref{fig:1}(a), $\hat{T}$ can be decomposed to 
\begin{equation}
    \hat{T}=\hat{V} \hat{L} \hat{U},
\end{equation}
where $\hat{V}$ and $\hat{U}$ are unitary matrices, and $\hat{L}$ is a real-valued diagonal matrix. For a passive system with no gain as our demonstration, without loss of generality, we can assume the form $\hat{L}=\mathrm{diag}[1,\eta]$ where $0\leq\eta\leq1$. The transformation $\hat{T}$ is unitary for $\eta=1$, and is otherwise non-unitary. 
The specific $\hat{T}$ that we aim to implement is a $50/50$ lossy beam-splitter transformation. To do so, we choose the following $2\times 2$ unitaries
\begin{gather}\label{eq:uv}
    \hat{U}=   \frac{1}{\sqrt{2}}
    \begin{bmatrix}
   1  & i \\
    i     & 1
   \end{bmatrix}, \ 
   \hat{V}=   \frac{1}{\sqrt{2}}
    \begin{bmatrix}
   i  & i \\
    -1     & 1
   \end{bmatrix}.
\end{gather}
Hence the resulting $\hat{T}$ is given by
\begin{gather}\label{eq:T}
    \hat{T}=   \frac{1}{2}
    \begin{bmatrix}
   -\eta+i  & -1+i\eta \\
    -1+i\eta     & \eta-i
   \end{bmatrix}, 
\end{gather}
where one can see that such a $\hat{T}$ always exhibits a $50/50$ splitting ratio for arbitrary values of $\eta$.

Such a linear transformation can be achieved using an extra/auxiliary port in photonic circuits. Consider a $3\times 3$ matrix $\hat{M}$, to obtain the desired $\hat{T}$, we apply a the following decomposition
\begin{equation}\label{eq:M}
    \hat{M}=\hat{M}_V \hat{M}_L \hat{M}_U,
\end{equation}
where
\begin{gather}
    \hat{M}_U=   
    \begin{bmatrix}
   \hat{U}  & 0 \\
   0            & 1
   \end{bmatrix},\ 
   \hat{M}_V=   
    \begin{bmatrix}
   \hat{V}  & 0 \\
   0            & 1
   \end{bmatrix},\
 \hat{M}_L
 =
  \begin{bmatrix}
   1 & 0 & 0\\
   0 & \cos \theta & i\sin\theta\\
   0 & i\sin \theta & \cos \theta
   \end{bmatrix},
   \label{eq:MUVL}
\end{gather}
with $\cos \theta \equiv \eta$ ($0\leq \theta\leq \pi/2$).
We implement $\hat{M}$ using photonic circuits as sketched in Fig.~\ref{fig:1}(b). While the implementation of a general $2\times 2$ unitary needs a Mach-Zehnder interferometer (MZI) that includes two beam splitters and two phase shifters, here for the specific $\hat{U}$ and $\hat{V}$ that we have chosen, the configuration can be simplified: $\hat{U}$ can be achieved using a $50/50$ beam splitter between waveguides 1 and 2. Together with a direct transmission of waveguide 3, we obtain $\hat{M}_U$. $\hat{V}$ is implemented by a $\pi/2$ phase shifter in waveguide 1 followed by a $50/50$ beam splitter of waveguides 1 and 2; Combining the direct transmission in waveguide 3 we therefore have $\hat{M}_V$. Between $\hat{M}_U$ and $\hat{M}_V$ we implement an additional beam splitter between waveguide 2 and the auxiliary mode, waveguide 3.  This realizes $\hat{M}_L$, where waveguide 3 plays the role of the loss channel, and the splitting ratio between modes 2 and 3 is proportional to the loss (1-$\eta$) of \eqref{eq:T}.
Inserting $\hat{U}$ and $\hat{V}$ given in \eqref{eq:uv} into \eqref{eq:M} and \eqref{eq:MUVL}, we find
\begin{gather}
     \hat{M}
 =
 \begin{bmatrix}
 \hat{T} & \begin{matrix} -\frac{\sin \theta}{\sqrt{2}}  \\ i\frac{\sin \theta}{\sqrt{2}}  \end{matrix} \\  \begin{matrix} -\frac{\sin \theta}{\sqrt{2}} & i\frac{\sin \theta}{\sqrt{2}}  \end{matrix}   & \cos \theta
\end{bmatrix}.
   \label{eq:M2}
\end{gather}
While the system has three input ports and three output ports and performs a unitary transformation $\hat{M}$, if we only look at the sub-system of input and output ports 1 and 2, we obtain the desired non-unitary transformation $\hat{T}$ given in \eqref{eq:T}.

Now we formulate two-photon interferences in such photonic circuits described by transformation $\hat{M}$. We note that $\hat{M}$ governs the transformation of photon creation operators of photons between the input and output of this system. Specifically,
\begin{equation}\label{eq:AdMBd}
    \hat{A^\dagger}=\hat{M}\hat{B^\dagger},
\end{equation}
where $\hat{A^\dagger}=[\mathbf{a^\dagger_1},\mathbf{a^\dagger_2}, \mathbf{a^\dagger_3}]^\mathrm{T}$ denotes the creation operators of photons $\mathbf{a^\dagger_n}$ at each input port indexed by $n$, and $\hat{B^\dagger}=[\mathbf{b^\dagger_1},\mathbf{b^\dagger_2}, \mathbf{b^\dagger_3}]^\mathrm{T}$ corresponds to the creation operators $\mathbf{b^\dagger_n}$ at the output ports. 
 
Using \eqref{eq:AdMBd} and by inserting \eqref{eq:M2}, we have
\begin{gather}\label{eq:a1d}
    \mathbf{a^\dagger_1}=\frac{1}{2}\left[(-\cos \theta +i) \mathbf{b^\dagger_1} +(-1+i\cos \theta)\mathbf{b^\dagger_2} -\sqrt{2}\sin\theta \mathbf{b^\dagger_3}\right],\\
    \mathbf{a^\dagger_2}=\frac{1}{2}\left[(-1+i\cos \theta) \mathbf{b^\dagger_1} +(\cos \theta-i)\mathbf{b^\dagger_2} +i\sqrt{2}\sin\theta \mathbf{b^\dagger_3}\right]. \label{eq:a2d}
\end{gather}

For our experiments, we prepare a two-photon input state described by the density matrix 
\begin{equation} \label{eq:rho}
\begin{split}
    &\hat{\rho}= \frac{1}{2} \left(\mathbf{a^\dagger_{1,s}} \mathbf{a^\dagger_{2,i}} \ket{0}\bra{0} \mathbf{a_{2,i}} \mathbf{a_{1,s}} + \mathbf{a^\dagger_{2,s}} \mathbf{a^\dagger_{1,i}} \ket{0}\bra{0} \mathbf{a_{1,i}} \mathbf{a_{2,s}} \right)\\
    &+\frac{e^{-\frac{\tau^2}{T^2}}}{2}  \left(\mathbf{a^\dagger_{1,s}} \mathbf{a^\dagger_{2,i}} \ket{0}\bra{0} \mathbf{a_{1,i}} \mathbf{a_{2,s}} +
    \mathbf{a^\dagger_{2,s}} \mathbf{a^\dagger_{1,i}} \ket{0}\bra{0} \mathbf{a_{2,i}} \mathbf{a_{1,s}}\right),
\end{split}
\end{equation}
where we label the photons using subscripts $s$ and $i$ to distinguish signal and idler photons generated from spontaneous parametric down conversion (SPDC), respectively. The generated photons are sent to a delay line generating a time-delay $\tau$ between the two photons. This gives rise to the $\exp(-\tau^2/T^2)$ coefficient in the second term of \eqref{eq:rho},
where we have assumed that the second-order coherence between photons in ports 1 and 2 can be described by a Gaussian function of the time-delay $\tau$~\cite{Hong1987}, and $T$ is essentially a coherence time.

Inserting \eqref{eq:a1d} and \eqref{eq:a2d} into \eqref{eq:rho}, we can calculate the nonlocal correlations between ports 1 and 2
\begin{equation}\label{eq:P12}
\begin{split}
    P_{1,2}&=\bra{0}\mathbf{b_{2,i}}\mathbf{b_{1,s}} \hat{\rho} \mathbf{b^\dagger_{1,s}}\mathbf{b^\dagger_{2,i}}\ket{0}+\bra{0}\mathbf{b_{1,i}}\mathbf{b_{2,s}} \rho \mathbf{b^\dagger_{2,s}}\mathbf{b^\dagger_{1,i}}\ket{0}\\
    &=\frac{1}{8}(\cos^2\theta+1)^2-\frac{1}{2}e^{-\frac{\tau^2}{T^2}}\cos^2\theta + \frac{1}{8}e^{-\frac{\tau^2}{T^2}}\sin^4\theta.
\end{split}
\end{equation}

\begin{figure}[t]
\centering
\includegraphics[width=\linewidth]{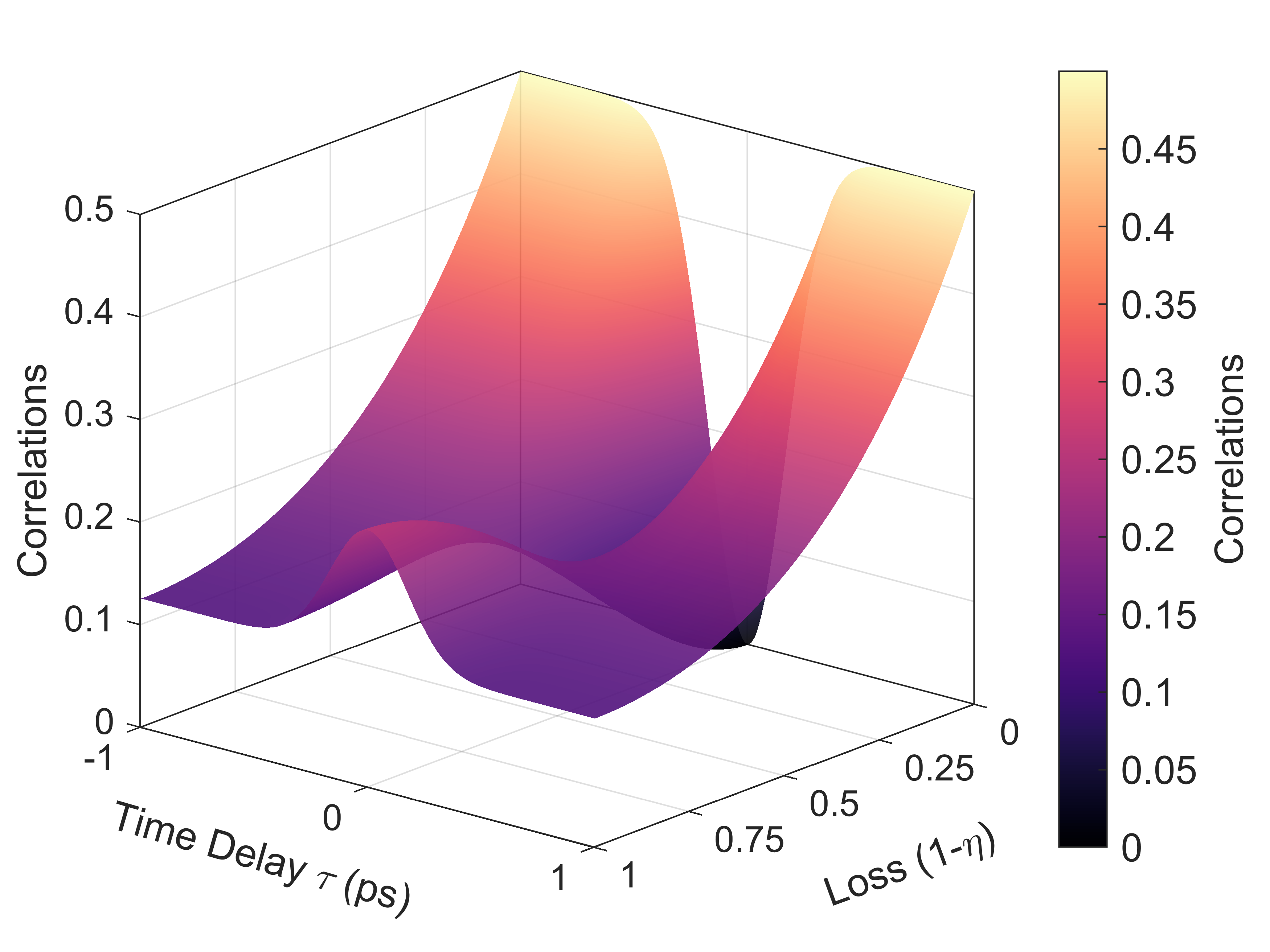}
\caption{Lossy HOM Photon Correlations. Two-photon correlations are plotted as a function of loss into an auxiliary mode. When loss equals 0, the quantum Hong-Ou-Mandel effect is seen with the visibility of 1; in the fully lossy case (Loss = 1) a peak in correlations is observed signifying a bunched state in the lower arm of the interferometer.}
\label{fig:2}
\end{figure}

In Fig.~\ref{fig:2} we plot the two-photon nonlocal correlation $P_{1,2}$ from output ports 1 and 2 predicted by \eqref{eq:P12} as a function of $\eta$ and $\tau$.
For the case without coupling to the loss channel, $\eta=1$, the $2\times2$ transformation $\hat{T}$ remains a unitary beam splitter, corresponding to the quintessential demonstration of quantum entanglement, a dip in correlations due to the Hong-Ou-Mandel (HOM) effect. When the photon pairs enter the interferometer with a long time delay $|\tau|$, the correlation between ports 1 and 2 is $P_{1,2} = 0.5$. When photon pairs enter the device at the same time ($\tau=0$, without time delay), photon pairs leave the device in bunched states due to their bosonic nature; thus, we observe a dip in correlations if one continuously varies $\tau$ ($P_{1,2} = 0$ at $\tau=0$). In contrast, in the case where $\hat{T}$ is a non-unitary matrix representing a lossy beam splitter ($\eta < 1$), qualitatively different behaviors take place in the two-photon interference. This is illustrated in Fig.~\ref{fig:2}. In the extreme case with $\eta = 0$ where $\hat{T}$ is a defective matrix describing a lossy beam splitter, one would see a peak instead of a dip if $\tau$ is continuously varied. Similar behaviors have been observed in other systems~\cite{Vest1373, Li2021}, but never in integrated photonic circuits.

To experimentally explore this effect, we implement the integrated waveguide structure as sketched in Fig.~\ref{fig:1}(b) using the direct femtosecond laser-written technique~\citep{Davis:96, Szameit_2010}, including the directional couplers and a phase shifter. The waveguides are inscribed in fused silica (Corning 7980). A commercial laser system (Coherent Mira/RegA) is used to create 150~fs pulses (800~nm wavelength, 445~nJ pulse energy, 100~kHz repetition rate), which are focused $200~ \mu \mathrm{m}$ below the silica surface using a standard objective ($20\times$ magnification, 0.36~NA). A three-axis Aerotech Inc. positioning system displaces the sample at 80~mm/min to write waveguides within the 10 cm fused silica sample. This technique results in elliptical mode profiles of $~9\times15~ \mu \mathrm{m^2}$ and features intrinsic losses $>0.3$~dB/cm at 815~nm. 50/50 directional couplers are attained by introducing a coupling region with $\sim19~\mu \mathrm{m}$ separation and a $\pi /2$ phase shift is introduced by altering the propagation constant in one waveguide by altering the mode profile with the faster writing speed of $85$~mm/min (highlighted in yellow Fig.~\ref{fig:1}(b)). The circuit is designed to be coupled directly with commercial v-groove fiber arrays by separating each waveguide by $127~ \mu \mathrm{m}$ at the end facet. Photons are detected using Excelitas avalanche photodetectors, and photon-pair correlations are recorded using a Becker–Hickl correlation card interfaced with LabVIEW with processing and simulations completed using MATLAB.

At the input of the device, we inject photon pairs generated through type I SPDC from a BiBO crystal, with one photon in each waveguide, i.e. a split two-photon state. The wavelength of the continuous-wave pump laser is 405~nm, and thus the generated photon pairs are 810 nm. Using a delay line, we can control the time delay $\tau$ between photons in a pair. Using an on chip 50/50 beam splitter (without auxiliary loss port, $\eta=1$) as reference we observe a HOM dip of $\xi=87\pm3\%$, where we define $\xi:=P_{1,2}(\tau=0)/P_{1,2}(\tau\rightarrow \infty)$. Such experimentally used photon source can be described by the density matrix
\begin{equation} \label{eq:rho_exp}
\begin{split}
    &\hat{\rho}^{exp}= \frac{1}{2} \left(\mathbf{a^\dagger_{1,s}} \mathbf{a^\dagger_{2,i}} \ket{0}\bra{0} \mathbf{a_{2,i}} \mathbf{a_{1,s}} + \mathbf{a^\dagger_{2,s}} \mathbf{a^\dagger_{1,i}} \ket{0}\bra{0} \mathbf{a_{1,i}} \mathbf{a_{2,s}} \right)\\
    &+\frac{\xi e^{-\frac{\tau^2}{T^2}}}{2}  \left(\mathbf{a^\dagger_{1,s}} 
    \mathbf{a^\dagger_{2,i}} \ket{0}\bra{0} \mathbf{a_{1,i}} \mathbf{a_{2,s}} 
    + \mathbf{a^\dagger_{2,s}} \mathbf{a^\dagger_{1,i}} \ket{0}\bra{0} \mathbf{a_{2,i}} \mathbf{a_{1,s}}\right),
\end{split}
\end{equation}
which differs with \eqref{eq:rho} only by a factor of $\xi$ in the latter terms. Since $\xi$ is only slightly smaller than unity, such an experimentally prepared state can allow us to see all the features predicted in the theory.

\begin{figure}[t]
\centering
\includegraphics[width=\linewidth]{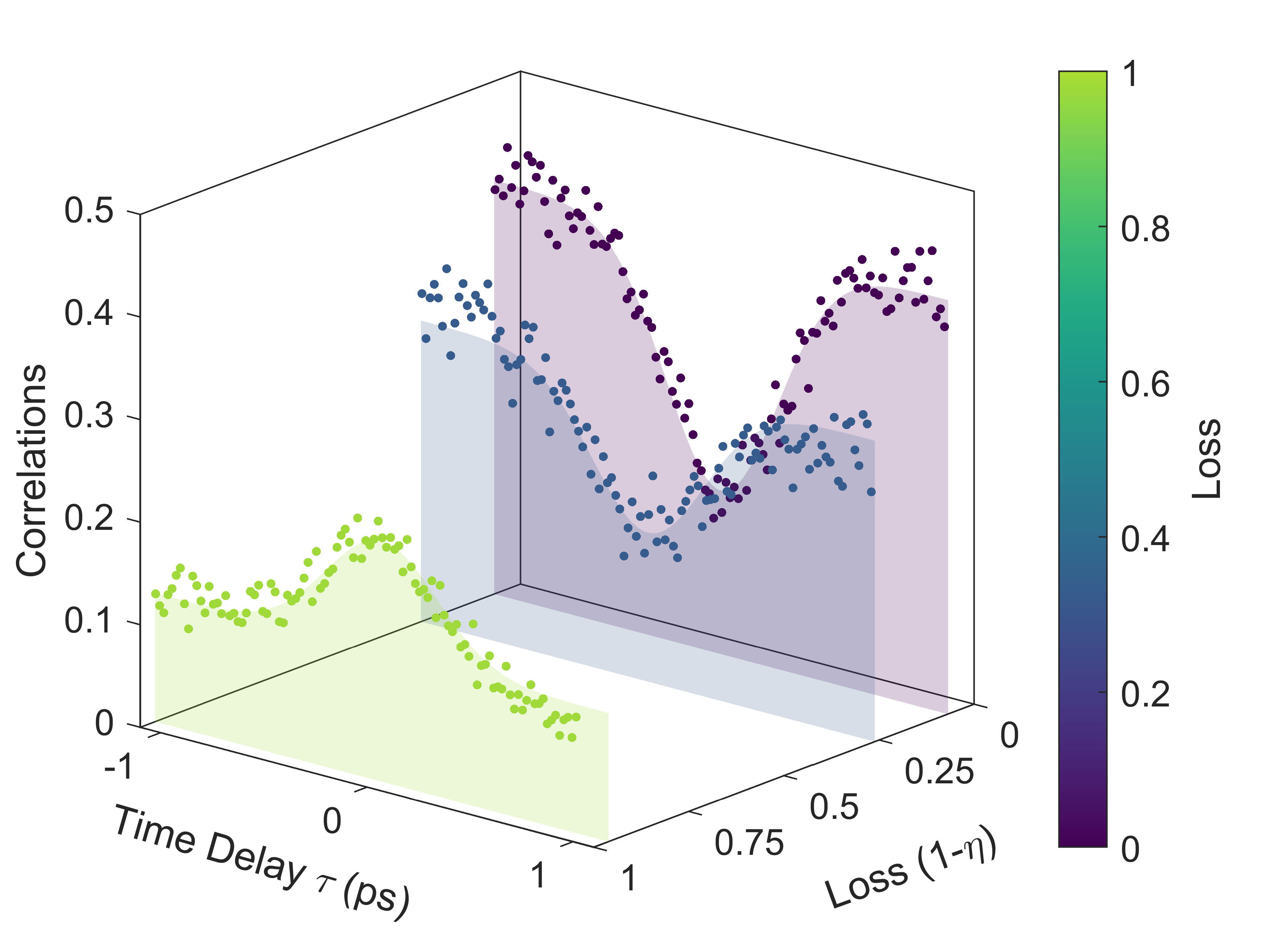}
\caption{Experimental two-photon coincidences between ports 1 and 2 in lossy MZI designed following the SVD approach. The correlations transform from anti-bunching to bunching as emulated loss increases.}
\label{fig:3}
\end{figure}

For our experiments, we fabricate an array of lossy MZIs with increasing coupling to the auxiliary mode. In Fig.~\ref{fig:3}, we plot the correlations for cases where the loss = 0.07, 0.26, and 0.96. For the case with the least loss (deep purple dots), we see the expected dip in correlations with a visibility of $60\pm3\%$ shown, close to the simulated visibility of 72\% for this loss (using a source visibility of $87\pm3\%$). As the loss increases, we see that the correlations for the indistinguishable state (with time delay $\tau = 0$) increase and the correlations for the distinguishable state (with time delay $|\tau| >$ 1~ps) decrease. This effect appears as the transition to a relative peak in correlations, which is due to the HOM interference at the first 50:50 beamsplitter and the differing probabilities of two photons entering the second beamsplitter. For the distinguishable case, at the first beamsplitter, there are four possible photon pathways, each with probabilities of 0.25 ($\ket{\psi} = \frac{1}{\sqrt{4}}\ket{0,0}+\frac{1}{\sqrt{4}}\ket{0,1}+\frac{1}{\sqrt{4}}\ket{1,0}+\frac{1}{\sqrt{4}}\ket{1,1}$). Only one of the above describes the case where two photons continue to the second beamsplitter ($\ket{1,1}$). While, for the indistinguishable case, HOM interference in the first beam splitter reduces these four possible cases down to two equally likely cases of $\ket{\psi} = \frac{1}{\sqrt{2}}\ket{0,0} \pm \frac{1}{\sqrt{2}}\ket{1,1}$, thus the final correlations after the second beamsplitter differ by a factor of two. As we can completely decouple photons via the auxiliary mode in the case of extreme loss, we observe a HOM peak with visibility of $87\pm7\%$, which perfectly matches the source visibility.

\begin{figure}[t]
\centering
\includegraphics[width=\linewidth]{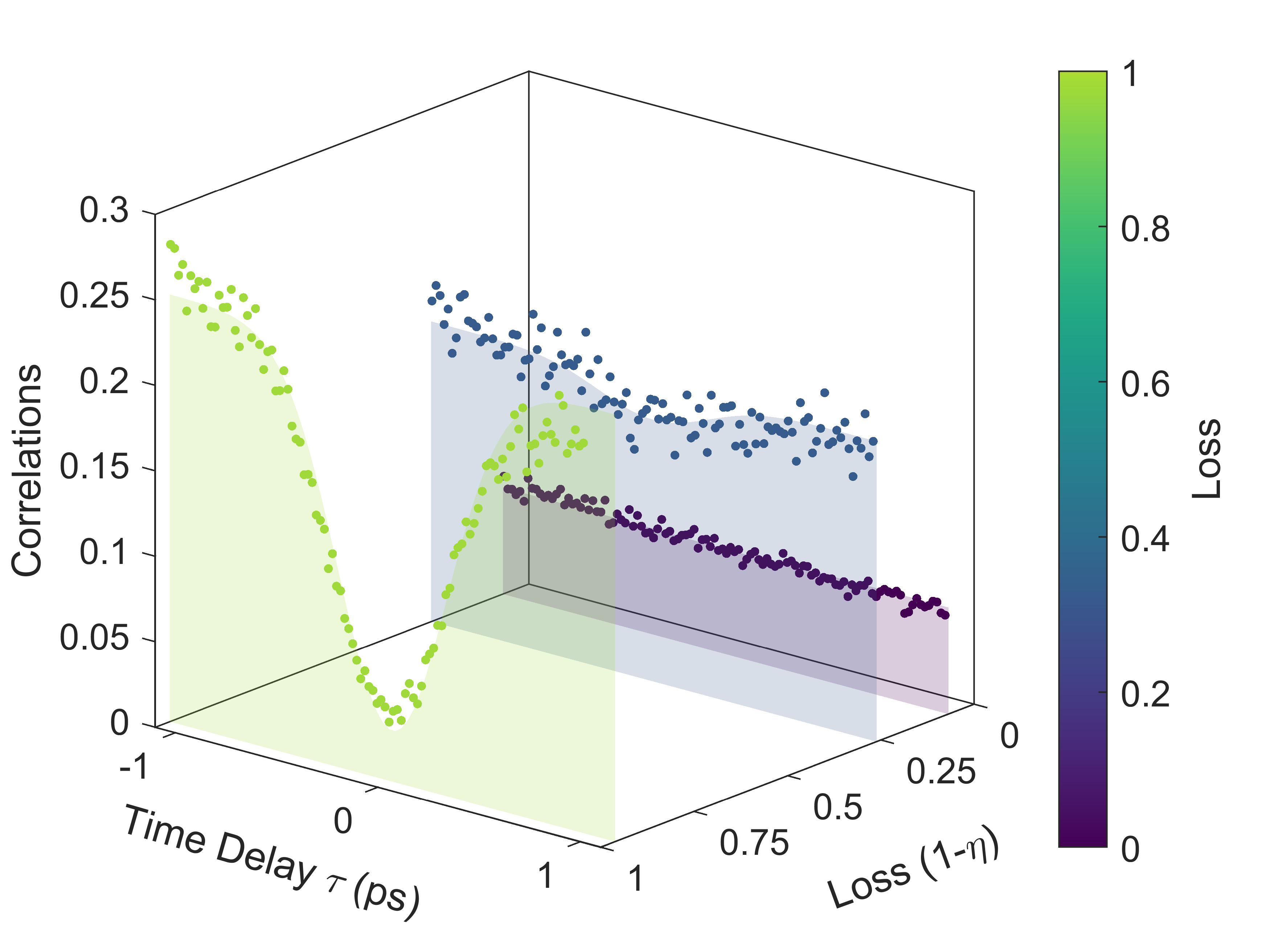}
\caption{Experimental two-photon coincidences with an auxiliary waveguide emulating the loss channel in MZI. The correlations show an increase in anti-bunching with an increase in emulated loss.}
\label{fig:4}
\end{figure}

An important feature emerging from this circuit design is that the presence of an auxiliary mode gives us access to measure the photons that are "lost" from the lossy beamsplitter, unlike a circuit where dissipation is induced through coupling to the bath. 
We can calculate the correlations between ports 1 and 3:
\begin{equation}
\begin{split}
    P_{1,3}&=\bra{0}\mathbf{b_{3,i}}\mathbf{b_{1,s}} \rho \mathbf{b^\dagger_{1,s}}\mathbf{b^\dagger_{3,i}}\ket{0}+\bra{0}\mathbf{b_{1,i}}\mathbf{b_{3,s}} \rho \mathbf{b^\dagger_{3,s}}\mathbf{b^\dagger_{1,i}}\ket{0}\\
    &=\frac{1}{4}(\cos^2\theta+1)\sin^2\theta- \frac{1}{4}e^{-\frac{\tau^2}{T^2}}\sin^4\theta.
\end{split}
\end{equation}
We make use of the feature and measure the correlations between the auxiliary mode and either channel of the MZI, which are normalized to 0.25 for long delay times and are plotted in Fig.~\ref{fig:4}. Here, it is important to note that the loss ($1-\eta$) refers to the loss from the MZI. We begin with the case without loss ($\eta = 1$), and no correlation events are observed. This is expected as the only correlation events detected are due to detector dark counts, which are Poissonian in time and show flat second-order correlation. As the coupling with the auxiliary mode (loss) increases, we see a dip in correlation events, up to a visibility of $88\pm3\%$. Here, we realize the most interesting facet of this implementation. By simply changing the measurement basis, that is, measuring correlations between the MZI outputs or measuring between the auxiliary mode and the MZI, we observe the inversion of the HOM peak to a HOM dip. We anticipate that this demonstration will shed light on photon interference under lossy transformations, help to develop an intuition for nontrivial correlations, and may offer insight for the design of larger-scale novel quantum photonic circuits.

In conclusion, we have demonstrated experimentally that photon-pair interference and the resulting correlations can be controlled precisely on a photonic chip by utilizing SVD, facilitating arbitrary linear transformations. We have shown that non-trivial correlations arise in the loss channel, which may be useful in the design of complex integrated photonic circuits. In the future, it may be of interest to expand these findings to more complex photonic circuitry and multi-photon states.

This work was supported by the Australian Research Council (DP190100277), Australian Research Council Centre of Excellence CE170100012, and the UA-DAAD exchange scheme. A.S. acknowledges funding from the Deutsche Forschungsgemeinschaft (grants SZ 276/9-2, SZ 276/19-1, SZ 276/20-1, SZ 276/21-1, SZ 276/27-1, GRK 2676/1-2023 ‘Imaging of Quantum Systems’, project no. 437567992, and SFB 1477 ‘Light–Matter Interactions at Interfaces’ (project no. 441234705). A.S. also acknowledges funding from the Krupp von Bohlen and Halbach Foundation as well as from the FET Open Grant EPIQUS (grant no. 899368) within the framework of the European H2020 programme for Excellent Science. 


%

\end{document}